\documentclass[twocolumn]{autart}    

\usepackage{graphicx,amsmath,amssymb,amsfonts,hyperref}

\begin{document}

\begin{frontmatter}

\title{Physics-Based Causal Lifting Linearization of Nonlinear Control Systems Underpinned by the Koopman Operator\thanksref{footnoteinfo}} 

\thanks[footnoteinfo]{This material is based upon work supported by National Science Foundation Grant NSF-CMMI 2021625. Corresponding author N.~S.~Selby.}

\author[eecs]{Nicholas Stearns Selby}\ead{nselby@mit.edu},    
\author[mech]{Filippos Edward Sotiropoulos}\ead{fes@mit.edu},    
\author[mech]{H. Harry Asada}\ead{asada@mit.edu}            

\address[eecs]{Department of Electrical Engineering and Computer Science, Massachusetts Institute of Technology, USA}  
\address[mech]{Department of Mechanical Engineering, Massachusetts Institute of Technology, USA}             


\begin{abstract}                          
Methods for constructing causal linear models from nonlinear dynamical systems through lifting linearization underpinned by Koopman operator and physical system modeling theory are presented. Outputs of a nonlinear control system, called observables, may be functions of state and input, $\phi(x,u)$. These input-dependent observables cannot be used for lifting the system because the state equations in the augmented space contain the time derivatives of input and are therefore anticausal. Here, the mechanism of creating anticausal observables is examined, and two methods for solving the causality problem in lifting linearization are presented. The first method is to replace anticausal observables by their integral variables $\phi^*$, and lift the dynamics with $\phi^*$, so that the time derivative of $\phi^*$ does not include the time derivative of input. The other method is to alter the original physical model by adding a small inertial element, or a small capacitive element, so that the system's causal relationship changes. These augmented dynamics alter the signal path from the input to the anticausal observable so that the observables are not dependent on inputs. Numerical simulations validate the effectiveness of the methods.
\end{abstract}

\end{frontmatter}

\section{Introduction}

Lifting linearization of nonlinear dynamical systems underpinned by the Koopman Operator theory has gained growing interest among the control, robotics, and other communities. Soft robotics \cite{soft-robot-koop}, human-robot interaction \cite{learn-koop-hri}, autonomous excavation \cite{koop-dig-1}, power systems \cite{koopman-app1}, and mission planning \cite{koopman-app3}, as well as many other robotics and control systems fields continue to demonstrate the efficacy of lifting linearizations for the modeling of control systems.

It is highly advantageous to use measured observables as opposed to computing synthetic nonlinear observables from state data. 

One of the major advantages of the Koopman-based lifting linearization is that it fits the framework of data-driven system identification. Nowadays, copious data are available at a lower cost in many application areas. In Dynamic Mode Decomposition (DMD), not only independent state variables but also dependent variables that can be measured directly from a nonlinear process are used for lifting the system. However, for control systems, one cannot use arbitrary observables measured from a dynamical system driven by exogenous inputs for lifting the dynamics. Observables may be functions of input $u$ as well as state $x$: $\phi(x, u)$. If any observables are input-dependent, the state equation in the lifted space will contain the time derivative of input and, thereby, the lifted system will not be causal. This anticausal observable problem is unique to dynamical systems with control and is a fundamental question when applying lifting linearization to control systems.

In the literature of Koopman operator theory, DMD, and lifting linearization, in general, the problem of input-dependent observables has been treated in a few different manners:
\begin{itemize}
\item The first method is to assume a state feedback controller, $u=u(x)$, and embed it into anticausal observables, $\phi(x, u) =\phi(x, u(x))$. This reduces the input-dependent observables to regular observables, functions of state variables alone \cite{abraham2019active,learn-koop,koop-w-ctrl}. Assuming a specific state feedback is restrictive and exogenous inputs cannot be involved in this formulation.
\item The second technique assumes that control $u$ is linearly involved in all anticausal observable functions:
\begin{equation}
    \phi(x, u) = \phi^*(x) + Du
    \label{eq:linear-filter}
\end{equation}
where $D$ is a constant matrix to be tuned to data, and $\phi^*(x)$ is a function of state alone. This is a practical approximation, but yields a significant error when the input nonlinearity is prominent \cite{koopman-app1,asada-dfl,l3}.
\item The third method is to assume the effect of the control input on the observables is quasi-periodic in time. This assumption allows singular components of the dynamic mode spectrum to be causal, and is thus primarily used as a theoretical stepping stone for modeling autonomous, periodic dynamical systems with no exogenous input \cite{mezic2016koopman}.
\item Finally, the causality problem does not occur if only causal observables are used for lifting \cite{koop-w-ctrl}. In case physically measured observables are used, however, one must be careful not to include those observables having exogenous inputs. Typically, causal observables are synthesized by artificially creating nonlinear functions of independent state variables alone. This prohibits the use of physically measured observables.
\end{itemize}

It is desired that the causality problem be solved without those restrictive assumptions. In particular, the assumption of equation (\ref{eq:linear-filter}), the linear involvement of exogenous inputs, does not hold for many nonlinear control systems, yielding significant errors. Furthermore, use of synthetic observables alone, although free from causality issues, has two major drawbacks. One is the lack of access to key signals that are often acquired through actual measurements of a nonlinear system. Use of broad physical measurements is much more effective than creating synthetic observables alone. The other drawback is noise. 

The standard procedure of the Koopman-based lifting linearization uses measured data of state variables and then augments the dataset with a large set of observables created from the data of state variables. In practice, measured state variables are corrupted with noise, resulting in significantly biased estimates of the linear model. Although the measurement noise of individual state variables is assumed to be uncorrelated to each other, the nonlinear transformations due to observables, which are nonlinear functions of state variables, cause complex correlation between the true state variables and the noise-corrupted terms of the observables, leading to significant errors in the resulting linear model. In contrast, the bias is small and amenable if both state variables and observables are measured directly. As detailed in Appendix~\ref{sec:noise}, the difference is clear and significant.

In this current work, we aim to analyze where anticausal observables for lifting arise in physical control systems, and solve the causality problem at the origin. Causality has a deep root in physics. We will consider physical control systems governed by basic physical laws, and present two methods to remove the causality problem.

The remainder of this article is structured as follows: in Section \ref{sec:prelim}, we introduce the basic concepts of causality in lifting linearizations and dual-faceted linearization. Our new approach is then introduced in Sections \ref{sec:admdc} and \ref{sec:idmdc}, and results are shown in Section \ref{sec:results} for a series of nonlinear models with anticausal variables. We conclude with a short summary and possible future work in Section \ref{sec:conc}.
\begin{figure}
    \centering
    \includegraphics[width=\linewidth]{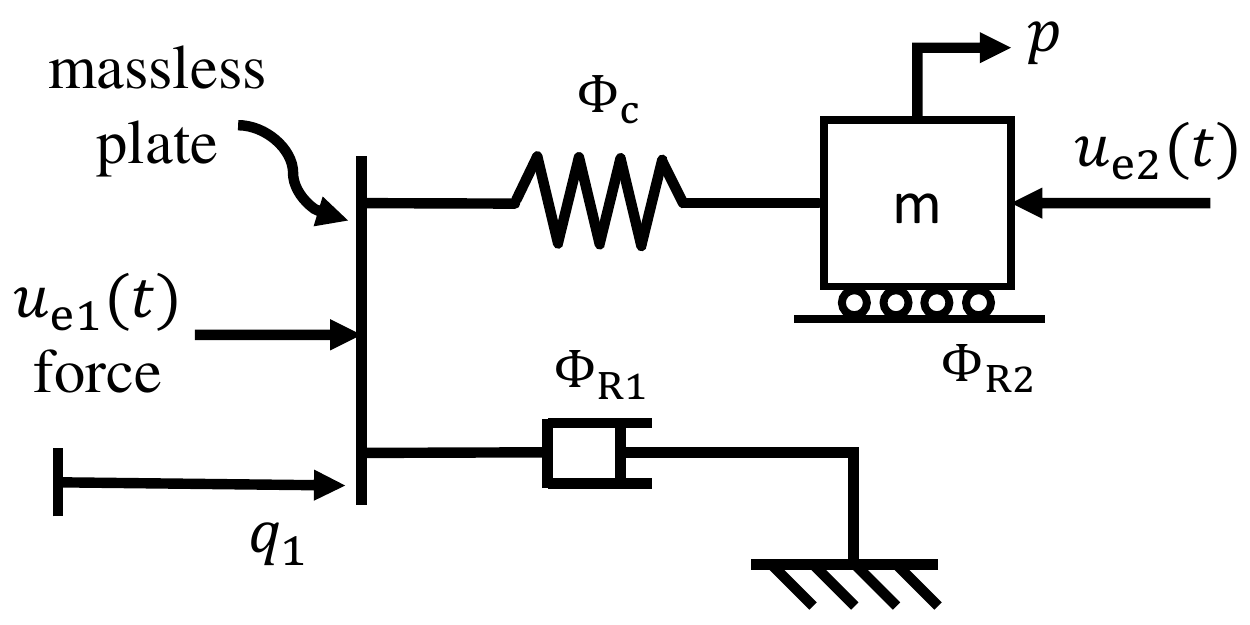}
    \caption{Example mass-spring-damper system.}
    \label{fig:prelim-msd}
\end{figure}

\section{Causality in Lifting Physical Control Systems}
\label{sec:prelim}

This section provides background information about causality of physical control systems and how anticausal problems occur when lifting dynamical systems.
\subsection{Causal Paths in Dynamical Systems}
In a physical control system, plant dynamics are governed by physical laws. Consider a lumped-parameter system consisting of basic elements, such as masses, springs, and dampers, subject to exogenous inputs as shown in Fig.~\ref{fig:prelim-msd}. The spring and two dampers are nonlinear, and their characteristics are represented with nonlinear functions called constitutive laws. The spring's constitutive law is given by $\mathrm{e}=\Phi_\mathrm{C}(q)$ where $q$ is elongation, or displacement, of the spring; e is the force generated by the spring; and $\Phi_\mathrm{C}(q)$ is a nonlinear, differentiable function. The constitutive law of each damper is given by a nonlinear, differentiable, and invertible function: $\mathrm{e}_{\mathrm{R}i} = \Phi_{\mathrm{R}i}(\mathrm{f}_i),\ i = 1,2$,
where $\mathrm{e}_{\mathrm{R}i}$ is the force generated by the $i^\mathrm{th}$ damper and $\mathrm{f}_i$ is the velocity of the damper\footnote{A damper is an energy-dissipative element in the mechanical domain. In the electric domain, the energy-dissipative element is a resistor with a constitutive law described in terms of voltage and current. In the fluid domain, it is described in terms of pressure and flow rate. Generically, we use effort e for force, voltage, and pressure and flow f for velocity, current, and flow rate.}. For the first damper, $\mathrm{f}_1 = dq/dt$.

The equation of motion of the mechanical system in Fig.~\ref{fig:prelim-msd} is given by $dp/dt = \mathrm{e_I}$ where $p$ is momentum of the mass and $\mathrm{e_I}$ is the resultant force acting on the mass given by $\mathrm{e_I} = \mathrm{e}-\mathrm{e_{R2}}+u_{\mathrm{e}2}$ where $u_{\mathrm{e}2}$ is the exogenous input force acting on the mass. From the icon model in Fig.~\ref{fig:prelim-msd}, we can also find the following conditions describing how these elements are connected to each other: $u_{\mathrm{e}1} = \mathrm{e}+\mathrm{e_{R1}}$ and $\mathrm{f_C} = \mathrm{f}_1 - \mathrm{f}_2$. Note that the momentum $p$ associated with the mass and the displacement $q$ associated with the spring can be used as independent state variables. 

\begin{figure}
    \centering
    \includegraphics[width=\linewidth]{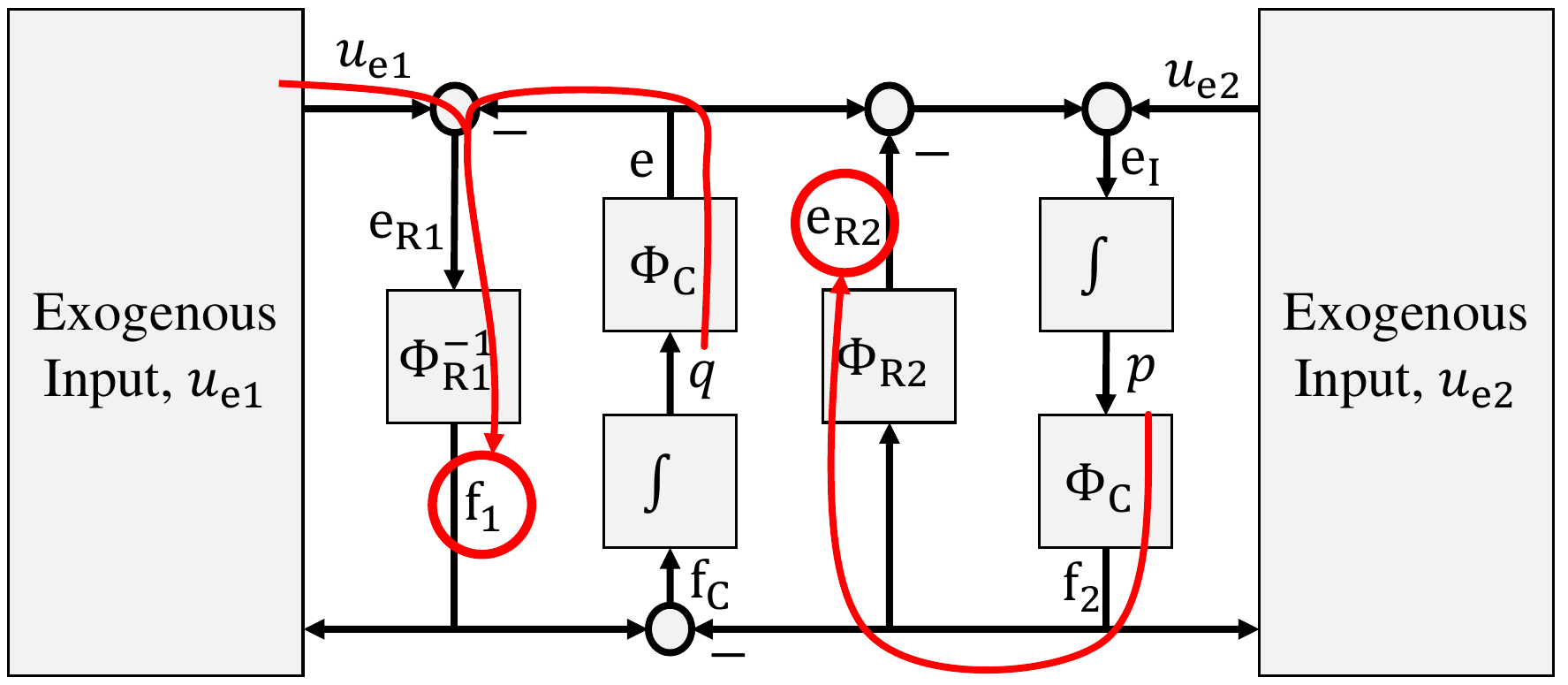}
    \caption{Element block diagram of nonlinear mass-spring-damper system from Fig.~\ref{fig:prelim-msd}.}
    \label{fig:prelim-blocks}
\end{figure}

This physical system is driven by exogenous input forces, $u_{\mathrm{e}1}$ and $u_{\mathrm{e}2}$, which propagate through the network of the elements and result in a change in the state variables $p$ and $q$. The propagation of signals can be graphically expressed with an element block diagram, as shown in Fig.~\ref{fig:prelim-blocks}. Note that the block of the mass has a specific direction of signals, that is, from force $\mathrm{e_I}$ to momentum $p$ and then to velocity $\mathrm{f}_2$. This direction cannot be reversed. As first addressed in Newton's Principia \cite{principia}, a resultant force acting on a mass determines the motion of the mass, and not the other way around. Along this causal direction, the momentum is determined by integration of the force, and then the velocity is determined by the algebraic relation, $\mathrm{f}_2 = p/\mathrm{m}$, the constitutive law of the mass. If the direction is reversed, i.e. velocity input and force output, a time derivative is involved in the signal transmission. In a sense, derivative uses future information and is not causal. Along the same line, the spring block in Fig.~\ref{fig:prelim-blocks} has input velocity $\mathrm{f_C}$ and output force e. If the input-output order is inverted, the element block includes a derivative. Therefore, the causal path is from velocity (flow) to force (effort). The block diagram in Fig.~\ref{fig:prelim-blocks} shows causal paths of all the signals originated in the two exogenous inputs \cite{bond-graph-textbook}. These signal directions also provide a computational procedure that can update state variables, $p$ and $q$, in response to exogenous inputs through integration and algebraic evaluation.

A dynamical system having this causal signal propagation is referred to as a system of integral causality \cite{bond-graph-textbook}. We assume that the dynamical systems considered in the current work are of integral causality. Integral causality can be checked conveniently with use of a bond graph. The causality propagation rules of bond graphs can determine causal input-output relations of individual elements in such a way that no conflict\footnote{Connecting the two outputs of two elements is a contradiction if both elements declare two outputs independently.} and no derivative occur in connecting all the elements. If this is possible, then the system is of integral causality. For readability of the paper, the following argument will be made without use of the bond graph notation. Instead, element block diagrams will be used for manifesting causal paths.

\subsection{Dual-Faceted Linearization (DFL)}
Dual-Faceted Linearization (DFL) \cite{asada-dfl} is a lifting linearization method based on a physical model of integral causality. It exploits knowledge of the connectivity of elements, or network structure, as illustrated in the icon model in Fig.~\ref{fig:prelim-msd}. Often we do not know the nonlinear function of each element's constitutive law, but we know how the mass, spring, and damper are connected to each other. In DFL, knowledge of constitutive laws is not required; experimental data and element connectivity information allow for lifting linearization. 

In DFL, the output variable of each nonlinear element is called an auxiliary variable. Given $n_\mathrm{a}$ nonlinear elements involved in a dynamical system, auxiliary variables are collectively denoted as $\eta \in \Re ^{n_\mathrm{a}}$.  In the example of Fig.~\ref{fig:prelim-msd}, the auxiliary variables are $\eta = (\mathrm{f}_1, \mathrm{e}, \mathrm{e_{R2}})^\intercal$. The dynamic system is lifted by using these auxiliary variables to obtain a linear model:
\begin{equation}
    \begin{cases}
        \dot{x} =  \mathrm{A}_x x +  \mathrm{A}_{\eta} \eta +  \mathrm{B}_{x} u \\ 
        \dot{\eta}= \mathrm{H}_x x  + \mathrm{H}_{\eta}\eta +  \mathrm{H}_u u
    \end{cases}
    \label{eq:dfl-cont}
\end{equation}
where $ x \in \Re^n$, $ u \in \Re^r$, and $\mathrm{A}_x, \mathrm{A}_{\eta}, \mathrm{B}_{x}, \mathrm{H}_x, \mathrm{H}_{\eta}$, and $\mathrm{H}_u$ are constant matrices with consistent dimensions. The first equation is exact and is determined by the element connectivity and linear elements, while the second equation is determined by regression. The approximation accuracy can be improved by further augmenting the space with observable functions of both independent state and auxiliary variables $\phi(x,\eta)$ \cite{igarashi2020mpc}, and the robustness of modeling is shown in \cite{igarashi2020robust}.

Note that causality plays a critical role in determining the output of each element. As discussed previously,  flow (e.g. velocity, current, flow rate, etc.) is the output for an inertial element, and effort (e.g. force, voltage, pressure, etc.) is the output of a capacitive element. On the other hand, the output of a resistive element depends on how the elements are connected to each other. Through causal path analysis, the direction of signal transmission is determined and, thereby, the output of a nonlinear resistive element is determined. 

Each of the auxiliary variables has a clear physical meaning and may be measured with sensors. For example, auxiliary variable e can be measured with a force sensor attached to the spring in Fig.~\ref{fig:prelim-msd}, and the output of the left damper, that is, the auxiliary variable $\mathrm{f}_1$ can be measured with a velocity sensor.

In physical systems, the connectivity of elements is basically linear if it is governed by  Kirchhoff's Voltage and Current Laws or the Generalized Kirchhoff's Loop and Node Rules \cite{kirchhoff}. In an electric circuit, all the voltages along a closed loop sum to zero, which is a linear relationship. In a mechanical system, all the forces at a mass, including an inertial force, sum to zero, which is also a linear relationship. The DFL formulation exploits this linearity in element connectivity to represent a state equation as a linear differential equation of independent state variables and auxiliary variables, which are outputs of all the nonlinear elements.

Nonlinearity comes from constitutive laws. Therefore, the output variables of nonlinear elements contain key properties of the nonlinear dynamical systems. It has been reported that lifting linearization using both independent state variables and auxiliary variables is more efficient for approximating a nonlinear dynamical system than doing so with independent state variables alone \cite{asada-dfl,jerry-dfl}. This implies that a richer collection of observables can be created from the augmented space of both state and auxiliary variables: $\phi = \phi(x, \eta)$.

A caveat in lifting a nonlinear dynamical system using auxiliary variables is that $\eta$ may be a function of exogenous inputs: $\eta = \eta(x, u)$.

\subsection{Anticausal Auxiliary Variables}
This section examines how auxiliary variables can be connected to exogenous inputs. First consider energy-storage elements, i.e. inertial and capacitive elements. For an inertial element, the output of the element is a flow variable. If the element's constitutive law $\Phi_\mathrm{I}$ is nonlinear, then the flow, f, is an auxiliary variable. Using this variable for lifting the system,
\begin{equation}
    \frac{d\mathrm{f}}{dt} = \frac{d\Phi_\mathrm{I}}{dp}\frac{dp}{dt}
\end{equation}
where $dp/dt$ is the time derivative of a state variable, that is, a state equation. Therefore, no time derivative of exogenous inputs is involved. The output of a capacitive element is an effort variable with time derivative
\begin{equation}
    \frac{d\mathrm{e}}{dt} = \frac{d\Phi_\mathrm{C}}{dq}\frac{dq}{dt}
\end{equation}
Again, no exogenous input can be involved.

An anticausal auxiliary variable may be generated only at energy-dissipative (resistive) elements. In the element block diagram in Fig.~\ref{fig:prelim-blocks}, two resistive elements are involved. Note that the input-output relations are different for the two resistive elements. For $\Phi_{\mathrm{R}1}^{-1}$, the input is effort $\mathrm{e}_{\mathrm{R}1}$ and the output is flow $\mathrm{f}_1$. For $\Phi_{\mathrm{R}2}$, the input is a flow and the output is an effort. These directions are dictated by the propagation of causal signals through the network. Auxiliary variables are determined based on this causality analysis. Namely, flow $\mathrm{f}_1$ and effort $\mathrm{e}_{\mathrm{R}2}$ are auxiliary variables. Because there exists a causal signal propagation path from exogenous input $u_{\mathrm{e}1}$ to auxiliary variable $\mathrm{f}_1$, this auxiliary variable is a function of the exogenous input: $\mathrm{f}_1=\Phi_{\mathrm{R}1}^{-1}(u_{\mathrm{e}1} -\Phi_\mathrm{C}(q))$. On the other hand, only state variable $p$ is involved in the causal path of the auxiliary variable $\mathrm{e}_{\mathrm{R}2}$. Therefore, this auxiliary variable is causal: $\mathrm{e}_{\mathrm{R}2}=\Phi_{\mathrm{R}2}(\Phi_\mathrm{I}(p))$. This auxiliary variable can be used as an observable for lifting the space.

\section{Augmented Lifting Linearization}
\label{sec:admdc}

\subsection{Resistors Connected to a Loop Junction}
We are interested in modeling systems that obey the generalized Kirchhoff loop and node rules \cite{kirchhoff} and contain resistive elements. As discussed in Section \ref{sec:prelim}, anticausal observables result only from energy-dissipative elements like resistors and dampers. Such elements can be connected to other elements via a loop or a node.

\begin{figure}
    \centering
    \includegraphics[width=\linewidth]{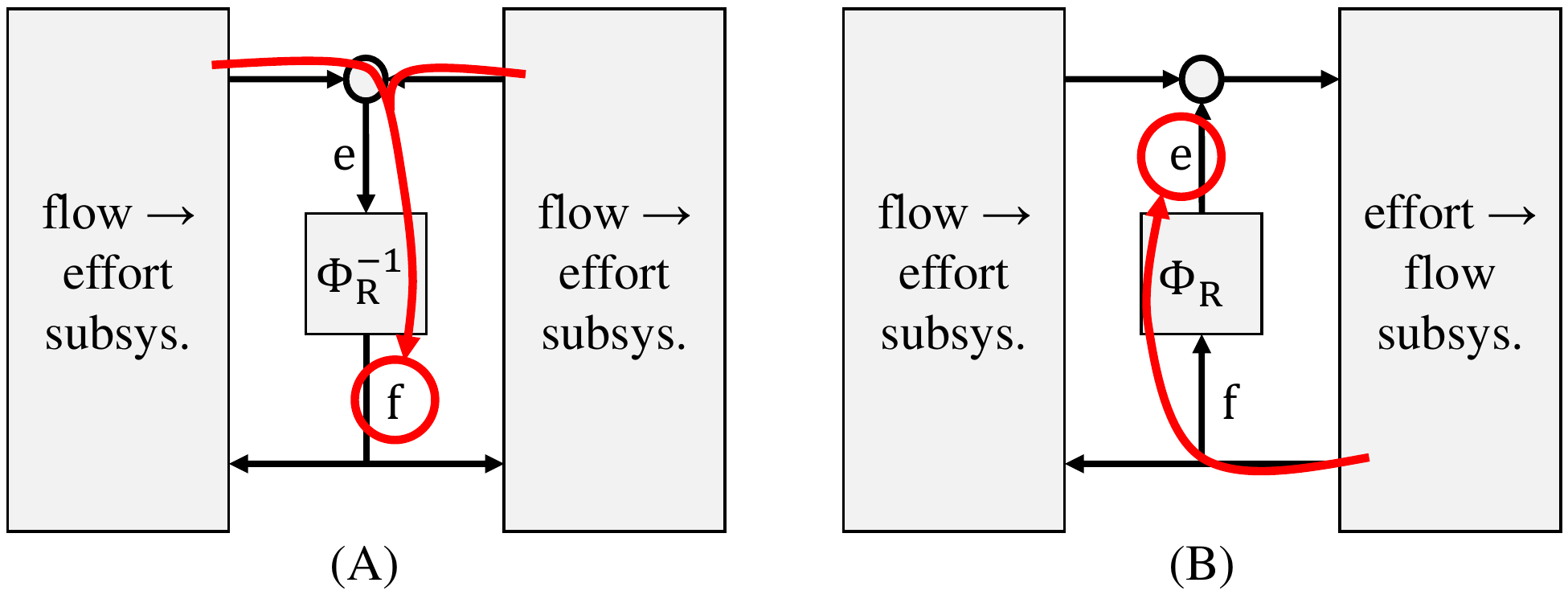}
    \caption{Two possible configurations of a resistor connected to a loop junction.}
    \label{fig:blocks-loop}
\end{figure}

Fig.~\ref{fig:blocks-loop} illustrates the two cases in which a resistor's connectivity is governed by Kirchhoff's Loop Rule: either the output of the resistor is a flow variable, as shown in Fig.~\ref{fig:blocks-loop}~(A), or the output of the resistor is an effort variable, as shown in Fig.~\ref{fig:blocks-loop}~(B). Note that the output e in Fig.~\ref{fig:blocks-loop}~(B) comes from the right subsystem, as indicated with a red path. This allows exogenous inputs from that subsystem to be involved in flow variable f. In Fig.~\ref{fig:blocks-loop}~(A), the output f comes from both subsystems, providing two paths from which to involve exogenous input.

\begin{figure}
    \centering
    \includegraphics[width=0.6\linewidth]{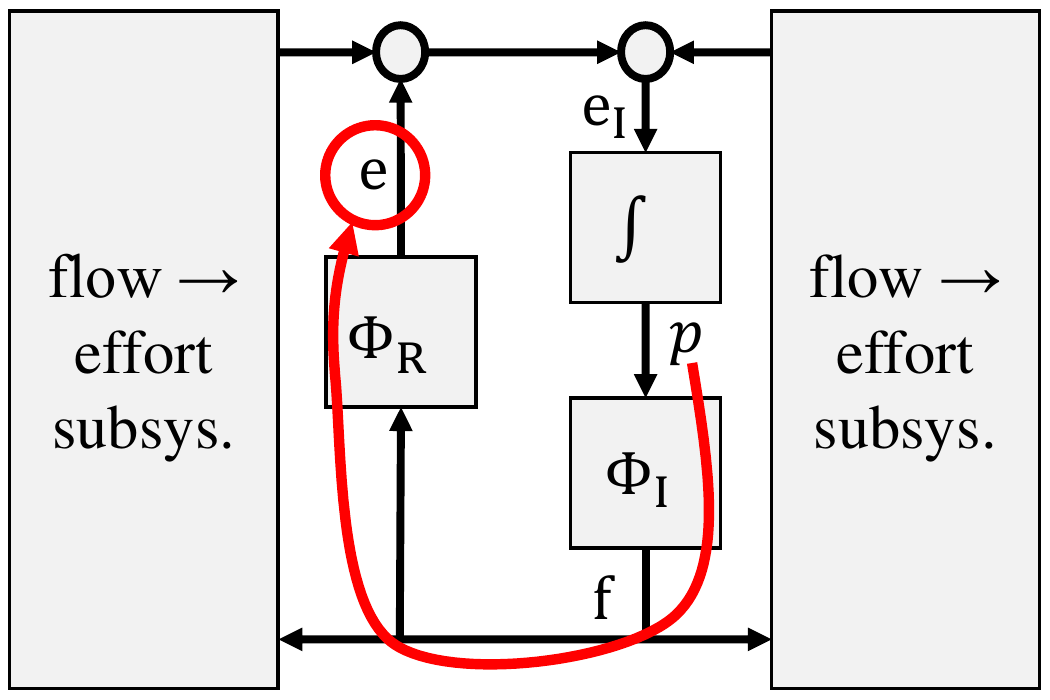}
    \caption{The system in Fig.~\ref{fig:blocks-loop} augmented with an inertial element. The inertial element produces output f in response to input $\mathrm{e_I}$ due to the inertia's inherent causality. This dictates the resistor to alter its input-output causal direction. As a result, the new output e is a nonlinear function of momentum $p$, which is an independent state variable, as shown with the red path from $p$ to e. By Kirchhoff's Loop Rule, a  flow variable is common to all the elements connected in the loop. Therefore, the output of the inertial element f is distributed to the resistor and the subsystems. In turn, the input $\mathrm{e_I}$ is determined such that all the efforts along the loop sum to zero.}
    \label{fig:blocks-loop-aug}
\end{figure}

Here, the objective is to intercept the causal paths connecting the subsystems to the resistor output, thereby preventing any exogenous input coming through the subsystems from being algebraically involved in the observable. This can be achieved by adding an inertial element connected to the loop junction. As illustrated in Fig.~\ref{fig:blocks-loop-aug}, the inertial element added to the system must output a flow variable f in response to the effort input $\mathrm{e_I}$ due to the causality, as addressed previously. 
The addition of the inertial element dictates that the output of the resistor is an effort. As a result, the new auxiliary variable e comes from the state of the inertia, i.e. momentum $p$, as indicated by the red path, and cannot be directly influenced by exogenous input from the subsystems. Therefore, the new auxiliary variable e can now be used as an observable to safely lift the state space.

\subsection{Resistors Connected to a Node Junction}
\begin{figure}[b]
    \centering
    \includegraphics[width=\linewidth]{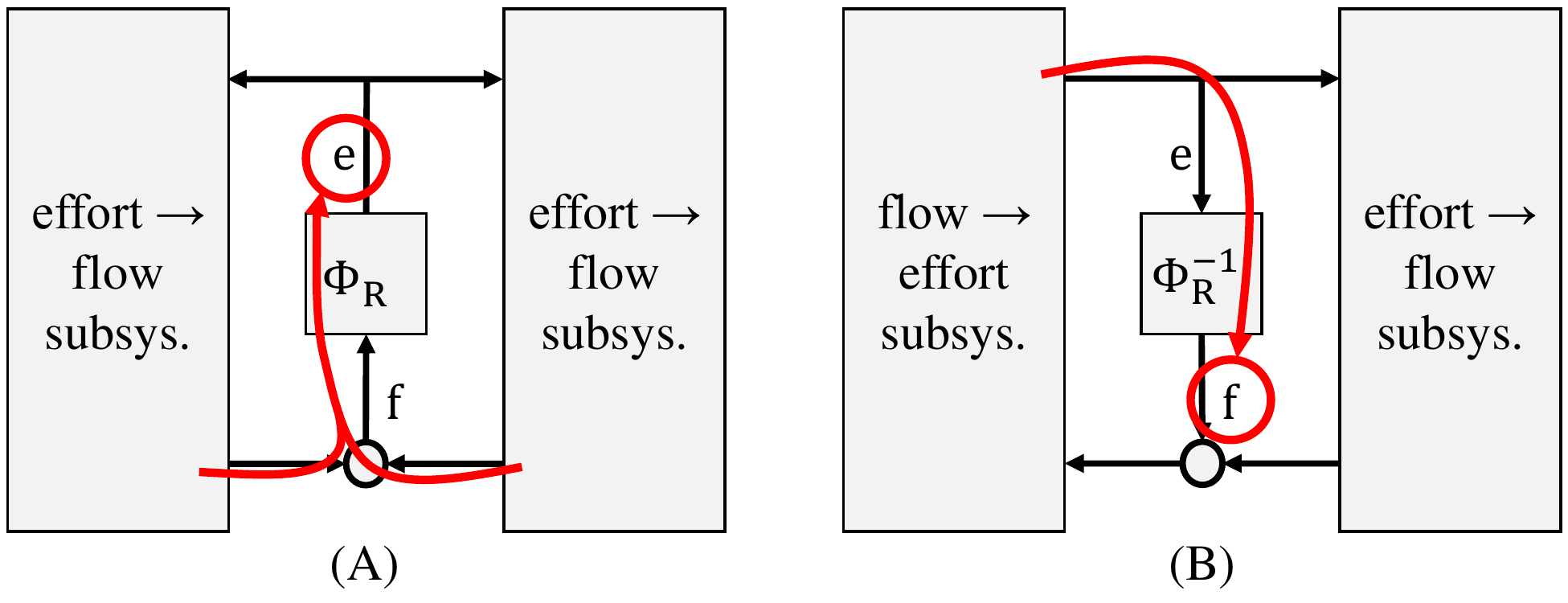}
    \caption{Two possible configurations of a resistor connected to a node junction.}
    \label{fig:blocks-node}
\end{figure}

\begin{figure}
    \centering
    \includegraphics[width=0.6\linewidth]{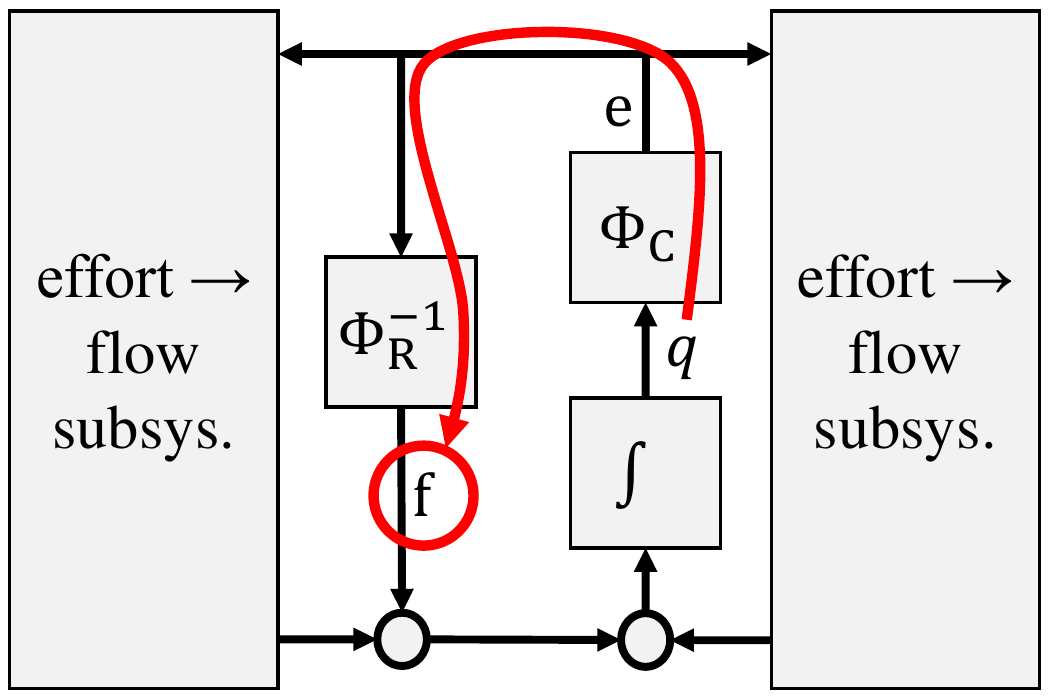}
    \caption{The system in Fig.~\ref{fig:blocks-node} augmented with a capacitive element to solve the causality problem.}
    \label{fig:blocks-node-aug}
\end{figure}

Similarly, Fig.~\ref{fig:blocks-node} illustrates the two cases in which a resistor's connectivity is governed by Kirchhoff's Node Rule, with Fig.~\ref{fig:blocks-node}~(A) showing the general case for effort-output and Fig.~\ref{fig:blocks-node}~(B) for flow output. These represent dual cases for the systems illustrated in Fig.~\ref{fig:blocks-loop}. By swapping the effort and flow variables in Fig.~\ref{fig:blocks-loop}, we can obtain the block diagrams in Fig.~\ref{fig:blocks-node}. Along the same line, by swapping effort and flow variables in Fig.~\ref{fig:blocks-loop-aug}, we can obtain a system with no causality problem. See Fig.~\ref{fig:blocks-node-aug}. Note that, in lieu of the inertial element, a capacitive element is added to the original system.

In all cases, causal paths are retained within the local elements consisting of the resistor and either an inertial or capacitive element added to the system. This prohibits the causal paths from adjacent subsystems that may contain exogenous inputs.

\subsection{Re-Routing Causal Paths}

The above method is summarized in the following Proposition.

\textbf{Proposition} Consider a dynamical system of integral causality where element connections are governed by the generalized Kirchhoff's loop and node rules. Causal paths from subsystem containing exogenous input can be prohibited from reaching the outputs of resistive elements by either adding an inertial element if the resistor is connected to a loop junction, or a capacitive element if the resistor is connected to a node junction. If a resistor is connected to a loop junction, the output of the augmented resistor is given by $\mathrm{e}=\Phi_\mathrm{R}\left( \Phi_\mathrm{I}(p)\right)$ where $\Phi_\mathrm{R}(\mathrm{f})$ and $\Phi_\mathrm{I}(p)$ are the constitutive laws of the resistor and added inertia, respectively. If the resistor is connected to a node junction, the output of the augmented resistor is given by $\mathrm{f}=\Phi_\mathrm{R}^{-1}\left( \Phi_\mathrm{C}(q)\right)$ where $\Phi_\mathrm{R}^{-1}(\mathrm{e})$ is the inverse constitutive law of the resistor and $\Phi_\mathrm{C}(q)$ is the constitutive law of the capacitor. These output variables are causal and can therefore be used to lift the dynamics.

\textbf{Remark 1.} The added inertial and capacitive elements can be linear with ``mass'' m and ``capacitance'' C such that $\mathrm{f}=p/\mathrm{m}$ and $\mathrm{e}=q/\mathrm{C}$. Because these are linear elements, the outputs of these elements are not auxiliary variables; no lifting is necessary for f and e. $\square$

\textbf{Remark 2.} Small values should be picked for the mass m and capacitance C so that the impact upon the dynamical system may be small. To select an initial value, one can use a Bode plot of the linearized system to determine the bandwidth of operation. The mass or capacitance can be selected so that it is small enough such that its breakpoint frequency is significantly higher than those of the other elements of the system. Thereby, the dynamics of the system can remain mostly unaffected in the low-frequency band useful for modeling. $\square$

In the icon model of Fig.~\ref{fig:prelim-msd}, the plate to which the spring and damper $\Phi_{\mathrm{R}1}$ are attached was assumed massless. This resulted in an anticausal auxiliary variable $\mathrm{f}_1$. Applying the proposition, we add a small inertia to resolve the anticausality problem, effectively modeling the plate in Fig.~\ref{fig:prelim-msd} with a small mass. It is a ``modeling decision'' whether the mass of the plate is considered or ignored. However, it has a significant impact on the causality of the lifted dynamics.

\textbf{Remark 3.} In the proposition, an energy-storage element, an inertia or a capacitor, is added to the loop or node junction to which the resistor of interest is connected. However, the energy-storage element can be added to other parts of the system so long as it prohibits the causal path from exogenous inputs from reaching the resistor. There may be multiple solutions to a given anticausal system. $\square$

\textbf{Remark 4.} If the output of a resistor is directly driven by an exogenous input, e.g. $\mathrm{e}=\Phi_\mathrm{R}(u_\mathrm{f})$ or $\mathrm{f}=\Phi_\mathrm{R}^{-1}(u_\mathrm{e})$, observations of e or f do not reflect any dynamics of the system. Therefore, these auxiliary variables can be omitted from the lifted model. $\square$

\section{Lifting with Integrated Auxiliary Variables}
\label{sec:idmdc}
As discussed above, lifting linearization through augmentation of the original system can solve the causality problem of auxiliary variables. However, the method inevitably alters the original dynamics, which, although slight, may not be allowed in some applications. This section introduces an alternative method that does not alter the dynamics, but creates a virtual instrument.

Recall that energy-storage elements do not incur the causality problem in lifting the system. Each energy-storage element contains an integrator that possesses an independent state variable. The time derivatives of these state variables are, by definition, given by $dp/dt = \mathrm{e}$ and $dq/dt = \mathrm{f}$ where effort and flow variables, f and e, respectively, can be expressed as functions of state variables, auxiliary variables, and exogenous inputs. Time derivatives of input are not involved in these expressions. This leads to causal state equations. 

Based on this observation, now consider the following ``trick'' to resolve the causality problem. Suppose that a virtual instrument measures the integral of the output of an energy-dissipative element. Let $\phi_{\mathrm{R}u}(x(t), u(t))$ be the output of a nonlinear, dissipative element that is anticausal for lifting. The virtual instrument produces $\phi^*(t)\triangleq \int_0^t \phi_{\mathrm{R}u}(x(\tau),u(\tau)) d\tau$. Suppose that we use this integrated output of the energy-dissipative element for lifting the dynamics. Namely, the time derivative of $\phi^*(t)$ is given by $d\phi^*/dt = \phi_{\mathrm{R}u} (x,u)$. Because no input derivative is involved, the dynamics are causal.

\begin{figure}
    \centering
    \includegraphics[width=\linewidth]{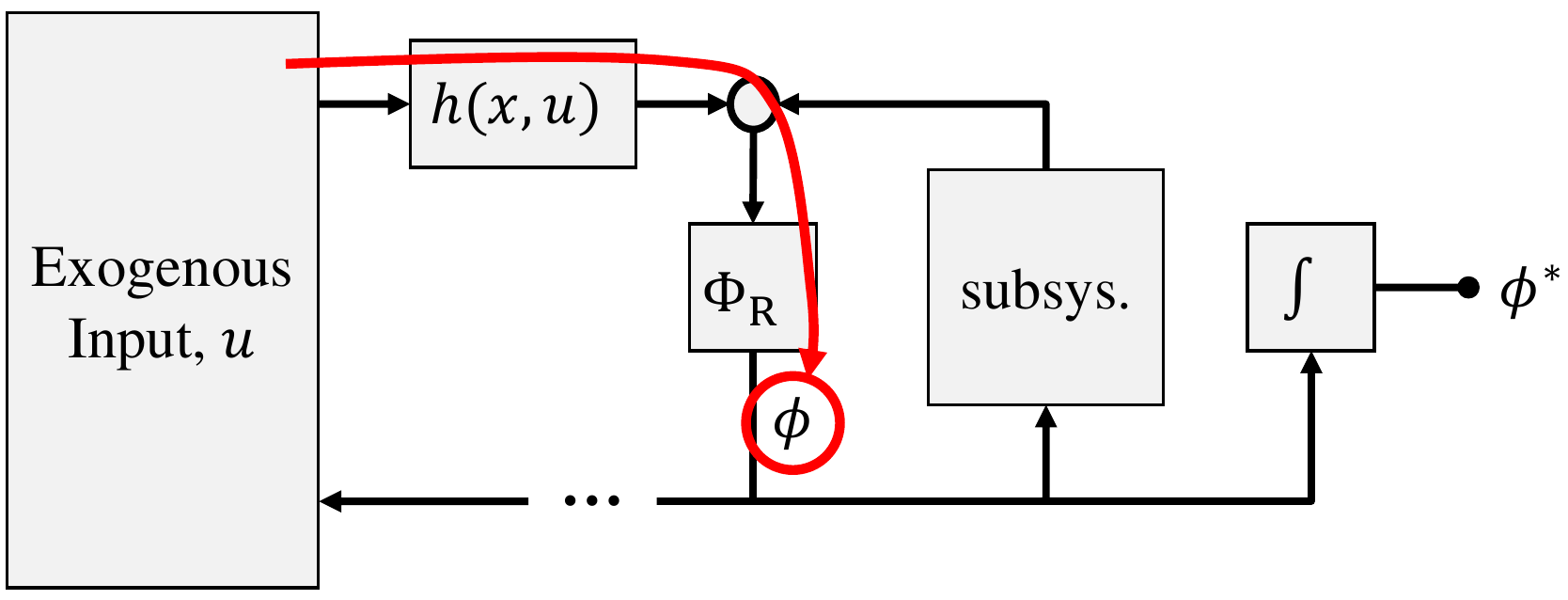}
    \caption{A dynamic system with an anticausal observable augmented with an integrator for causal lifting.}
    \label{fig:il2}
\end{figure}

This method, in a sense, treats the energy-dissipative element like an energy storage element, which possesses a state variable. Unlike the first method, where the original system is altered by adding a small inertial or capacitive element, this second method does not alter the dynamics. The integrator attached to the energy-dissipative element is only used for observation via a virtual instrument. As shown in Fig.~\ref{fig:il2}, the output of the virtual instrument $\phi^*$ does not influence the plant dynamics.




The above algorithm using Integrated observables for Lifting Linearization (IL2) is summarized below.

\textbf{IL2 Algorithm}
Use DFL to select $n_a$ auxiliary variables, $\phi(t)=\phi(x(t),u(t))$. Let $n_{\mathrm{R}u}$ be the number of nonlinear resistive elements in which $u(t)$ is involved. All the auxiliary variables associated with these input-dependent resistive elements are placed in a vector $\phi_{\mathrm{R}u}(x(t),u(t))\in \Re^{n_{\mathrm{R}u}}$. 

For each $\phi_{\mathrm{R}u,i}(x(t),u(t)),\ 1 \leq i \leq n_{\mathrm{R}u}$:
\begin{enumerate}
    \item Introduce a new variable:
    \begin{gather}
        \phi_{\mathrm{R}u,i}^*(t)\triangleq \int_0^t \phi_{\mathrm{R}u,i}(x(\tau),u(\tau)) d\tau
        \label{eq:idmdc}
    \end{gather}
    \item Remove $\phi_{\mathrm{R}u}(x(t),u(t))$ from the set of augmented state variables $\phi(t)$.
    \item If it is not already a state variable, append $\phi_{\mathrm{R}u,i}^*(t)$ to the set of state variables $x(t)$, and
    \item Append $\dot \phi_{\mathrm{R}u,i}^*(t)=\phi_{\mathrm{R}u,i}(x(t),u(t))$ to the state transition dynamics, $f$.
    \item Proceed to further lift the augmented state using an input-independent basis and/or regress a linear dynamic model to evolve the augmented state.
\end{enumerate}

\textbf{Remark 5} In some cases, $\phi_{\mathrm{R}u}^*(t)$ is already included in the state. If this happens, then the above algorithm instructs simply removing anticausal variables from the lifted state. $\square$

Consider a simple massless spring-damper system with state $x$ governed by the following dynamics:
\begin{equation}
    \dot x(t)=\mathrm{f}(t)\triangleq \Phi_{\mathrm{R}}^{-1}(u(t)-\mathrm{e_s}(t)),\quad \mathrm{e_s}(t)\triangleq \Phi_{\mathrm{C}}(x(t))
    \label{eq:old-toy}
\end{equation}
where $\Phi_{\mathrm{R}}$ and $\Phi_{\mathrm{C}}$ are nonlinear constitutive laws and the observables are selected using DFL to be $\phi(t)\triangleq \left(x(t),\mathrm{f}(t),\mathrm{e_s}(t)\right)^\intercal$. Note that the second observable, f, is a nonlinear function of exogenous input $u$, and therefore cannot be used for lifting. However, the integration of $\mathrm{f}(t)$ based on equation (\ref{eq:idmdc}) is the same as the state, $x$: $\mathrm{f}^*\triangleq \int \Phi_{\mathrm{R}}^{-1}(u(t)-\mathrm{e_s}(t))=x$. Therefore, the integrated auxiliary variable does not make any meaningful contribution. The IL2 algorithm instructs ignoring f altogether.

\section{Numerical Results}
\label{sec:results}
This section implements the augmented and integrated lifting linearization algorithms on a variety of simulated nonlinear dynamical systems with anticausal auxiliary variables and compares their performance against existing methods. 
The algorithms, written in SciPy \cite{scipy}, were computed on a laptop running Ubuntu 18.04.5 LTS. The codebase is hosted as a git repository at \cite{repo}. We used Adam's real-valued variable-coefficient ordinary differential equation solver \cite{vode} to perform the integration. We benchmark AL2 and IL2 against the following modeling algorithms:
\begin{itemize}
    \item \textbf{Dual-Faceted Linearization (DFL)} \cite{asada-dfl}, including the techniques to select observables using outputs of nonlinear elements connected to a lumped parameter model with integral causality. We also implement the linear anticausal filter to attempt to remove the effect of control input from observables.
    \item \textbf{Learned Lifting Linearization (L3)} \cite{l3} with a neural network with two hidden layers of 256 ReLU neurons trained in batches of 32 using an Adam optimizer \cite{adam} with $\alpha=10^{-5}$, $\beta_1=0.9$, $\beta_2=0.999$, and $\epsilon=10^{-8}$ until validation error begins to rise. We implemented the machine learning component of L3 in PyTorch \cite{pytorch}.
    \item \textbf{Koopman with Synthetic Observables of State (KSOS)} 
    All the observables are synthetically generated as nonlinear functions of state variables alone. No exogenous inputs are involved in the observables. Specifically, a polynomial basis of dimension 8 is used for generating data. 
    \item \textbf{Observables based on all Measured Quantities (OMQ)} 
    Emulating the DMD data-driven framework, data are generated based on the same physically meaningful quantities as DFL. All the observables are nonlinear functions of the measured variables, which may include anticausal observables. State and auxiliary variables are treated as physically-meaningful quantities, which can be measured, and a polynomial basis of order 8 is used for generating nonlinear observables. 
\end{itemize}

\subsection{Simple Numerical Example}

\begin{figure}
    \centering
    \includegraphics[width=\linewidth]{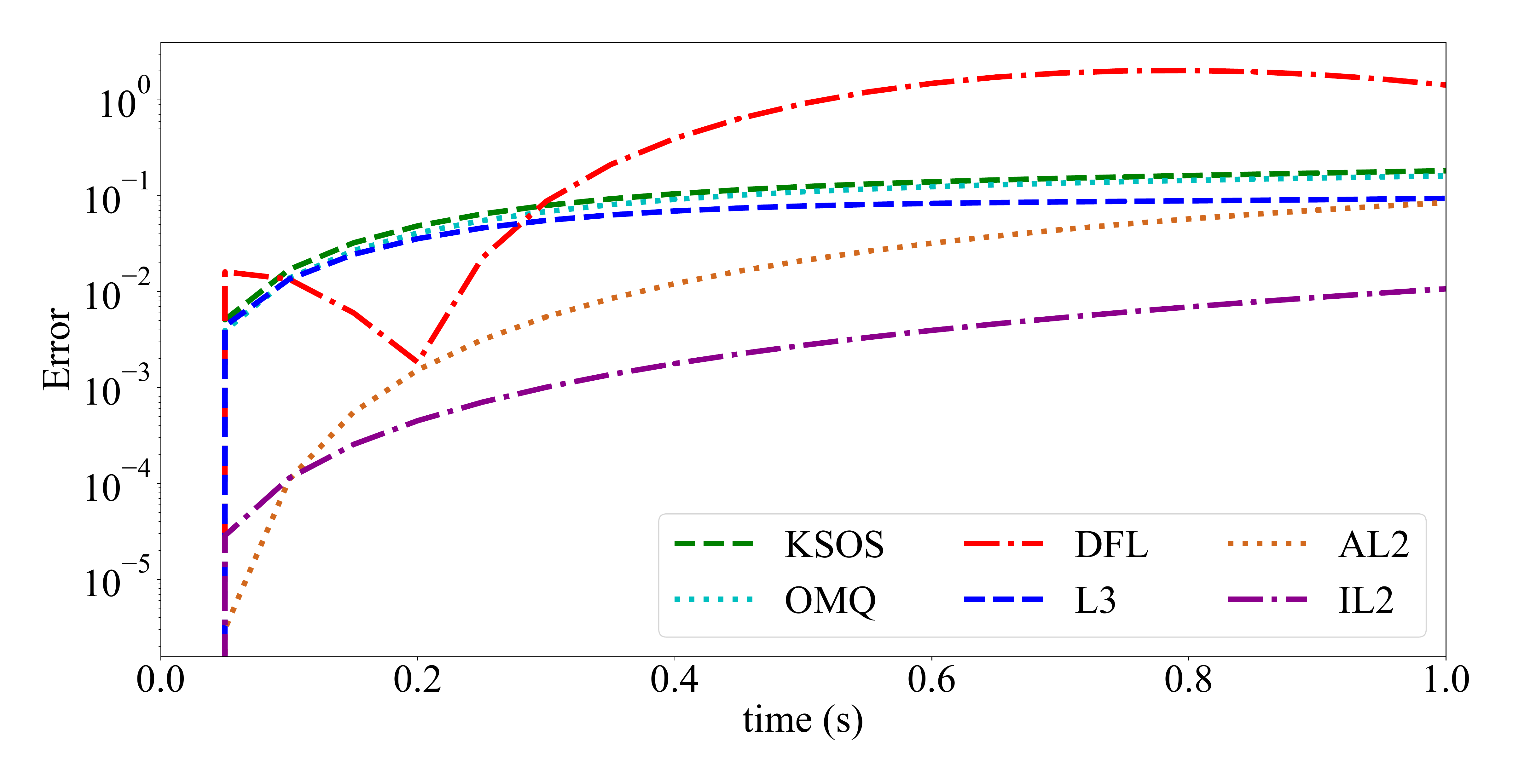}
    \caption{Simulation of numerical example from equation (\ref{eq:toy}) given a square wave control input with no noise. In the presence of moderate nonlinearity, AL2 and IL2 perform about as well as state-of-the-art techniques.}
    \label{fig:results-toy}
\end{figure}

Consider the following dynamical system with states $p$ and $q$ and control input $u$:
\begin{equation}
    \frac{d}{dt}\left( \begin{array}{c}
        p \\
        q
    \end{array} \right)=\left( \begin{array}{c}
        \Phi_{\mathrm{C}}(q) \\
        \Phi_{\mathrm{R}}\left(u-\Phi_{\mathrm{C}}(q)\right)-\Phi_{\mathrm{I}}(p)
    \end{array} \right)
    \label{eq:toy}
\end{equation}
where $\Phi_{\mathrm{C}}$, $\Phi_{\mathrm{R}}$, and $\Phi_{\mathrm{I}}$ are nonlinear constitutive laws associated with capacitive, resistive, and inertial elements in the system, respectively.

Based on DFL, we select the following observables: 
\begin{equation}
    \phi\triangleq \left(p, q, \Phi_{\mathrm{C}}(q), \Phi_{\mathrm{R}}\left(u-\Phi_{\mathrm{C}}(q)\right), \Phi_{\mathrm{I}}(p) \right)^\intercal
\end{equation}
Note that $\phi^{(4)}\triangleq \Phi_{\mathrm{R}}\left(u-\Phi_{\mathrm{C}}(q)\right)$ is a nonlinear function of control input and is therefore anticausal for lifting. For the implementation of AL2, we augment the system by adding a small mass $\mathrm{m}_0$ with a new state variable $p^*$ where $\dot p^*=u-\Phi_\mathrm{C}(q)-\Phi_\mathrm{R}(p^*/\mathrm{m}_0)$ and replace $\phi^{(4)}$ with a new observable $\mathrm{e_R}=\Phi_\mathrm{R}(p^*/\mathrm{m}_0)$. For the implementation of IL2, we replace $\phi^{(4)}$ with $\phi^*\triangleq \int \Phi_{\mathrm{R}}\left(u-\Phi_{\mathrm{C}}(q)\right)$ and augment equation (\ref{eq:toy}) with $\dot \phi^*=\Phi_{\mathrm{R}}\left(u-\Phi_{\mathrm{C}}(q)\right)$.

Assuming no noise in all measurements, we begin by simulating this system with the following constitutive laws: $\Phi_{\mathrm{C}}(q)=\mathrm{sgn}(q)q^2$, $\Phi_{\mathrm{R}}(\mathrm{e_R})=\mathrm{sgn}(\mathrm{e_R})\mathrm{e_R}^4$, and $\Phi_{\mathrm{I}}(p)=p^3$. The exogenous input given to the system is a square wave. The Sum of Squared Error (SSE) of each method is plotted in Fig.~\ref{fig:results-toy}. IL2 outperforms all other methods. In DFL and L3, it is assumed that the exogenous inputs are linearly involved in all the anticausal auxiliary variables, as shown in equation (\ref{eq:linear-filter}). Since this assumption does not hold due to the nonlinearity in $\Phi_{\mathrm{R}}$, the DFL and L3 models exhibit significant errors with SSEs of 270 and 17 times that of IL2, respectively. As in \cite{l3}, the OMQ model does not track the signal due to the inclusion of anticausal observable $\phi^{(4)}$ without first filtering out the dependence on $u$. The KSOS model, which did not include any auxiliary variables but instead lifted only with the state variables, achieved an SSE of 29 times that of IL2.

\begin{figure}
    \centering
    \includegraphics[width=\linewidth]{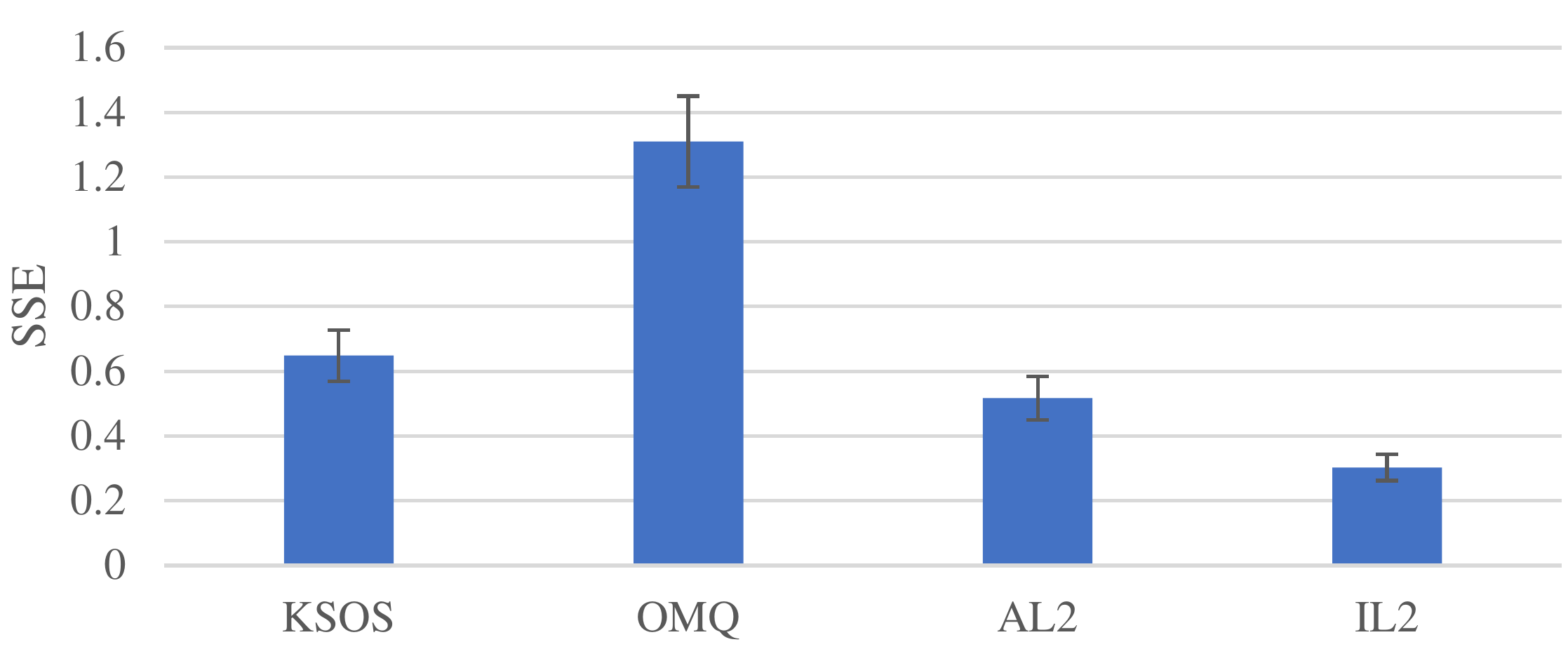}
    \caption{Average sums of squared error (SSE) for each algorithm modeling the same numerical example from equation (\ref{eq:toy}) with added nonlinearity given ten random control input signals and no noise. Standard error of each SSE is illustrated as error bars. The AL2 and IL2 model outperforms all other numerical models.}
    \label{fig:results-nonlin}
\end{figure}

\subsection{Increasing Nonlinearity}

To further examine the effect of system nonlinearity on the performance of AL2 and IL2, we repeat the same experiment using $\Phi_{\mathrm{I}}(p)=3p-3p^3$ and $\Phi_{\mathrm{R}}(\mathrm{e_R})=3\mathrm{e_R}^3-3\mathrm{e_R}$, reducing the coefficient of linear determination of both constitutive laws by 22\%. We repeat the experiment ten times using different random control input signals and plot the average sums of squared error for each method in Fig.~\ref{fig:results-nonlin}. The added nonlinearity breaks the underlying linearity assumption behind the anticausal filter, so the DFL and L3 models are omitted from this experiment. The OMQ model remains unable to track the system, and the KSOS model achieved an SSE double that of IL2.

\subsection{Comparison to Koopman}
KSOS shares fundamental similarities with AL2 and IL2. Neither KSOS nor AL2/IL2 lift the space with anticausal observables. 
However, the proposed algorithms have two key differences. First, IL2 uses the integral of an anticausal observable as a new causal observable for lifting the space. At the time of writing, the authors are not aware of any Koopman-based technique that uses an integral function, but it is precisely the use of integration that makes IL2 so effective. Second, AL2 and IL2 leverage DFL, which allows for the selection of informative observables so that the order of the linear dynamic model may be kept relatively small. 

\begin{figure}
    \centering
    \includegraphics[width=\linewidth]{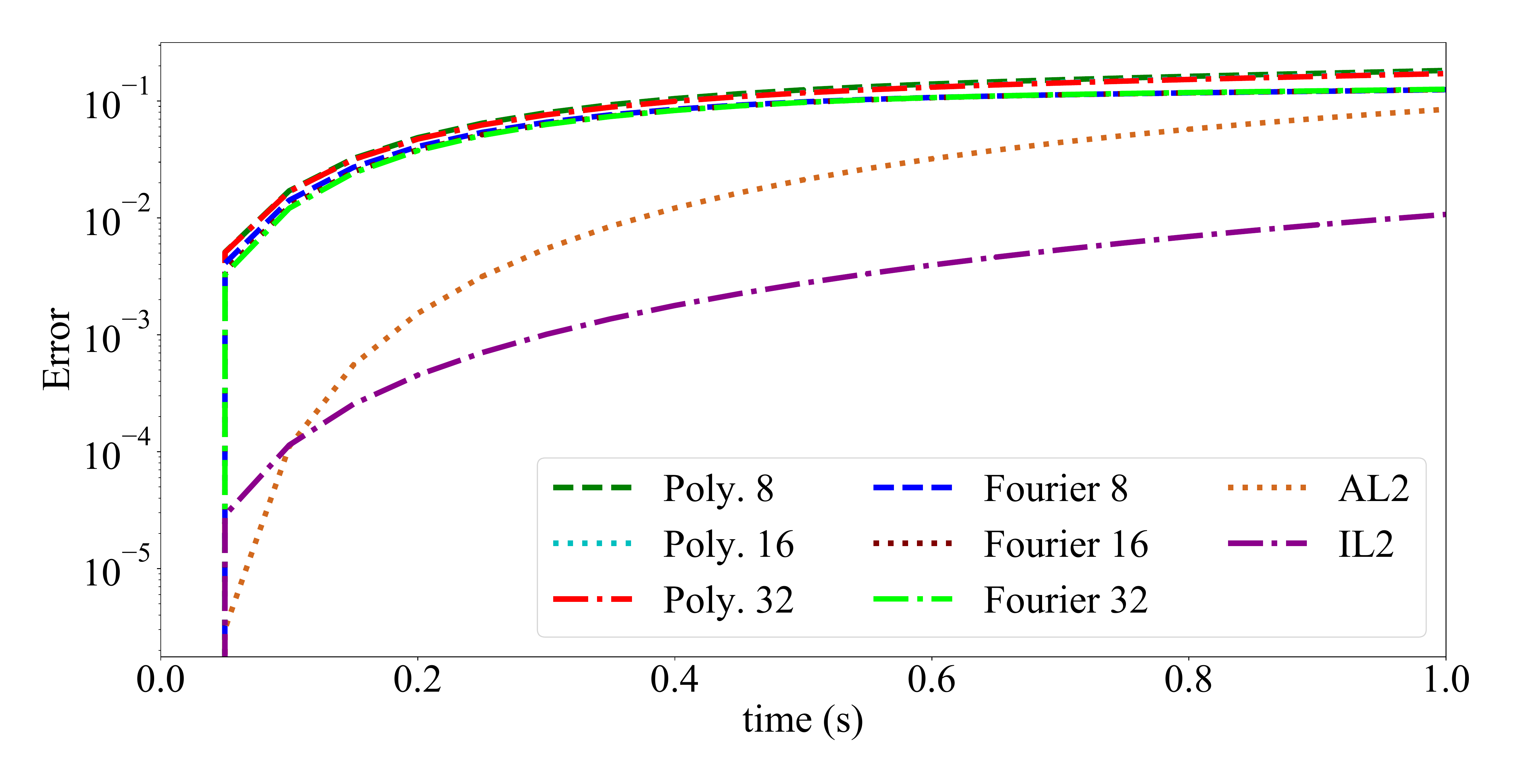}
    \caption{Comparison of AL2 and IL2 to KSOS with different basis functions.}
    \label{fig:results-bases}
\end{figure}

Fig.~\ref{fig:results-bases} illustrates the results of the experiment from Fig.~\ref{fig:results-toy} with six different implementations of KSOS: three with polynomial bases of dimension 8, 16, and 32, and three with Fourier bases \cite{fourier-basis} of dimension 8, 16, and 32. For comparison, the orders of the AL2 and IL2 models are only 7 and 6, respectively. As shown, improving the basis of KSOS without knowledge of the connectivity of elements does little to improve the results. This highlights the importance of accessing informative signals rather than manipulating basis functions alone.

\subsection{Noise in Observables}
\begin{figure}[b]
    \centering
    \includegraphics[width=\linewidth]{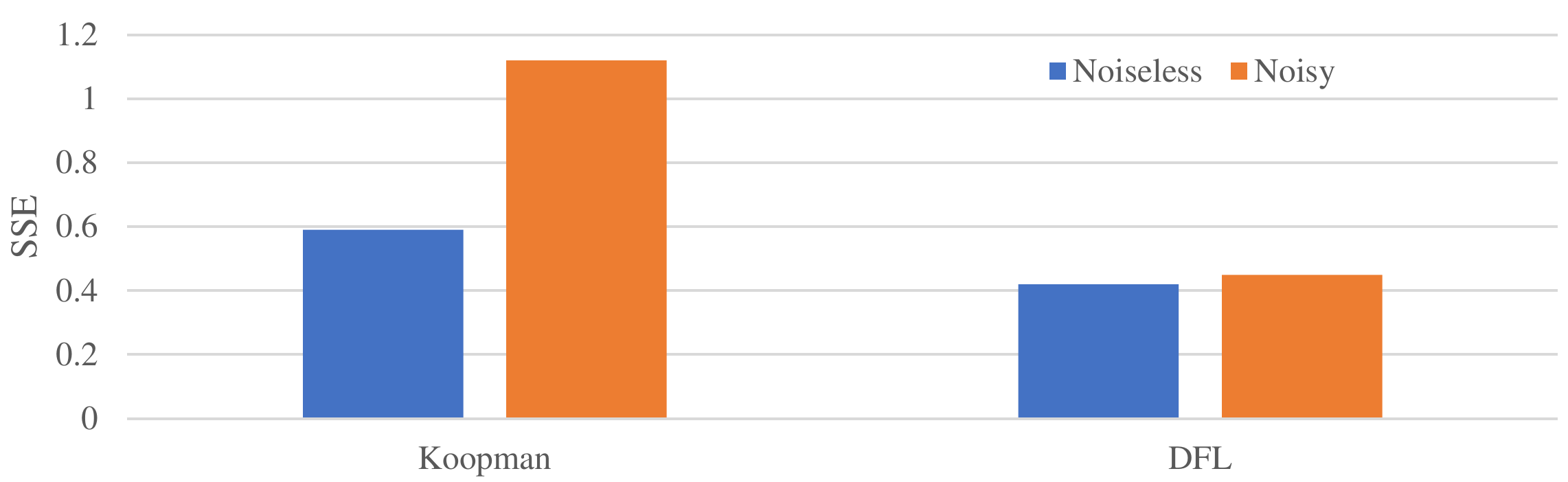}
    \caption{Comparison of model performance approximating the system from (\ref{eq:toy}) with and without noise. Because it uses synthetic observables, KSOS degrades much faster than DFL.}
    \label{fig:results-noise}
\end{figure}

Up to this point, all of the experiments illustrated have been noiseless. The KSOS algorithm, in particular, has benefited from the lack of noise in simulations. In this next experiment, we compare the noise characteristics of two approaches, DFL and KSOS. We repeat the experiment modeling the system in (\ref{eq:toy}) with a nonlinear constitutive law, $\Phi_\mathrm{R}(\mathrm{e_R})=1/(1+\mathrm{exp}(-4\mathrm{e_R}))-0.5)$. We repeat the experiment twice: once without noise, and once with zero-mean Gaussian noise with $\sigma=0.03$ added to each measurement, and compare the results in Fig.~\ref{fig:results-noise}. It is clear that DFL is considerably more robust in comparison to simply using the KSOS approach. This supports the argument in Appendix~\ref{sec:noise} showing that KSOS is more susceptible to noise as noisy signals are transmitted through nonlinear functions.

\subsection{Use Case for Inertial Augmentation}
\begin{figure}
    \centering
    \includegraphics[width=\linewidth]{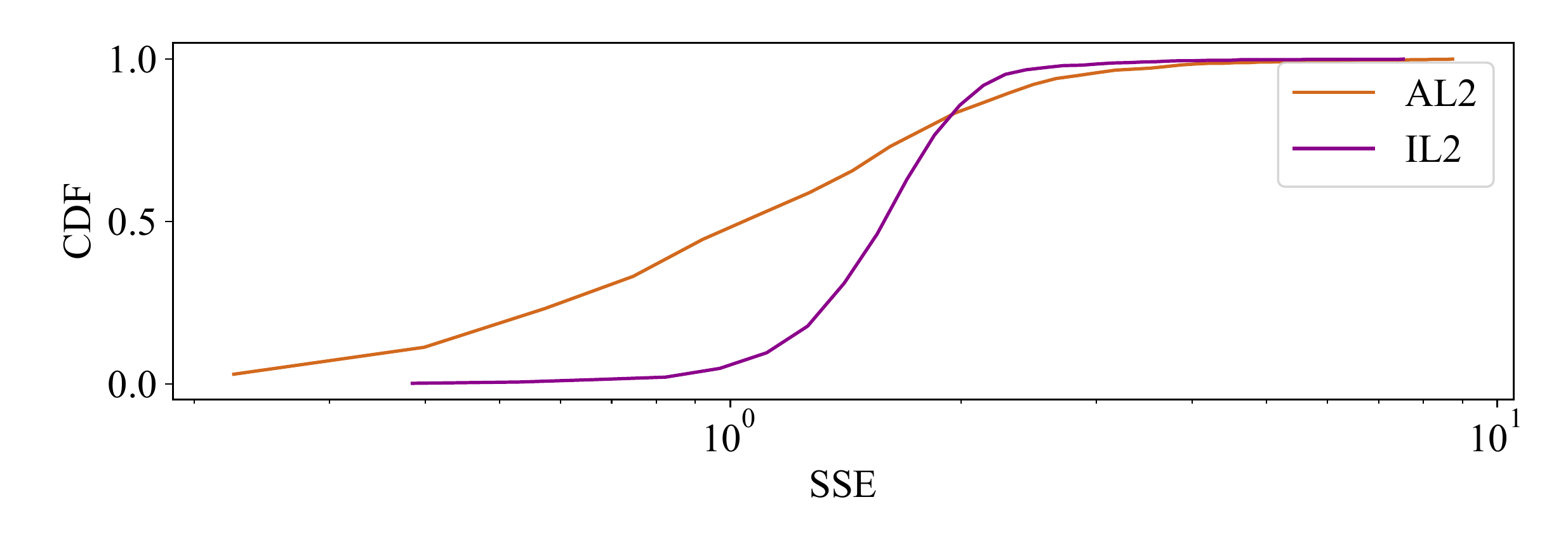}
    \caption{CDF of SSE for AL2 and IL2 models on the noiseless, massless, nonlinear spring-damper system defined by equation (\ref{eq:old-toy}). AL2 has a lower median error than IL2, but a greater 90$^{\mathrm{th}}$ percentile SSE.}
    \label{fig:results-admdc}
\end{figure}

In the above numerical examples, the integrated lifting linearization models have consistently outperformed the augmented lifting linearizations. Now we consider a case where the auxiliary variable cannot be replaced with its integral, because the integral is already an independent state variable. Let us revisit the situation in equation (\ref{eq:old-toy}), where the auxiliary variable $\mathrm{f}\triangleq \Phi_\mathrm{R}^{-1}(u-\Phi_\mathrm{C}(x))$ be replaced with $\phi^*\triangleq \int \mathrm{f}=x$, but $x$ is already an independent state variable. Therefore, IL2 instructs us to ignore important information about the dynamics of the system. In this case, AL2 can outperform IL2.

Because the performance of the modeling algorithms varies with the shape of the control input signal, we varied it. We simulate the massless spring-damper with constitutive equations $\Phi_\mathrm{R}^{-1}(\mathrm{e_R})=3\mathrm{e_R}(\mathrm{e_R}+1)(\mathrm{e_R}-1)$ and $\Phi_\mathrm{C}(x)=-\Phi_\mathrm{R}^{-1}(x)$ 1000 times with random control input signals and plot the Cumulative Distribution Function (CDF) of the SSE for each algorithm in Fig.~\ref{fig:results-admdc}. In most trials, the AL2 algorithm outperformed IL2 (median SSE of 1.0 and 1.6, respectively), but the IL2 algorithm had a more consistent performance with a lower 90$^\mathrm{th}$ percentile error of 2.1 compared to 2.4 for AL2.

\section{Conclusion}
\label{sec:conc}

Lifting a nonlinear control system using measured observables results in an anti-causal lifted model if exogenous inputs $u$ are involved in the observables $\phi(x,u)$. Based on physical modeling theory, this paper has addressed (a) how exogenous inputs propagate through a nonlinear dynamical system and can reach outputs of nonlinear elements, a class of observables causing the causality problem, and (b) how the causality problem can be solved. Observables that are outputs of nonlinear energy-storage elements never include exogenous inputs, while energy-dissipative elements may include exogenous inputs. Two methods have been presented for resolving the causality problem of energy dissipative elements.
The first method augments the system with either a linear inertial element or a linear capacitive element, thereby reversing the causality of the energy dissipative element and resolving the causality problem. The second method presented replaces input-dependent observables with integrals of linearly related causal observables, which also eliminates the causality problem inherent in predicting future values of an exogenous signal. Using these techniques, we synthesize a basis capable of lifting the state space of a nonlinear dynamic system into a more linear regime for modeling. The numerical results confirm that these techniques model nonlinear systems more accurately than any of the state-of-the-art algorithms against which we benchmarked. This work contributes to filling the gap between the Koopman-based lifting linearization of nonlinear autonomous systems and the lifted linear model of a nonlinear control system used for control design.

\bibliographystyle{plain}
\bibliography{root}

\appendix
\section{Noise in Observables}
\label{sec:noise}
As demonstrated by \cite{filippos}, using DFL to select observables achieves a considerable reduction in measurement noise compared to KSOS. Consider generic observables $\phi_t\triangleq \left( x_t^\intercal, \eta_t^\intercal \right)^\intercal$ where $\eta_t$ are observables that are not equivalent to states $x_t$ at time $t$. In DFL, $\eta_t$ is a vector of non-state measurements, and in KSOS, $\eta_t$ is a nonlinear basis applied to $x$. In real-world applications, physical measurements are corrupted by noise. Let $\tilde x_t\triangleq x_t+\epsilon_{x,t}$ and $\tilde \eta_t\triangleq \eta_t+\epsilon_{\eta,t}$ where $\epsilon_{x,t}$ and $\epsilon_{\eta,t}$ represent measurement noise affecting $x_t$ and $\eta_t$, respectively.

Consider the discrete-time version of equation (\ref{eq:dfl-cont}), the linear dynamic model learned using DFL:
\begin{equation}
    \left( \begin{array}{c}
        x_{t+1} \\
        \eta_{t+1}
    \end{array} \right) = A \left( \begin{array}{c}
        x_t \\
        \eta_t \\
        u_t
    \end{array} \right),\ A\triangleq \left( \begin{array}{ccc}
        A_x & A_\eta & B_x \\
        H_x & H_\eta & H_u
    \end{array} \right)
\end{equation}
When performing DFL, the state variables $x_t$ and auxiliary variables $\eta_t$ are physically measured with separate sensors. Therefore, the sensor noise in each of these two will be uncorrelated. Additionally, measurement noise is uncorrelated with measurements, i.e.
\begin{equation}
    \mathrm{E}\left[ \epsilon_{x,t} \epsilon_{\eta,\tau}^\intercal \right]=\mathrm{E}\left[ x_t\epsilon_{x,\tau}^\intercal\right]=\mathrm{E}\left[ \eta_t\epsilon_{\eta,\tau}^\intercal\right]=0\ \forall t,\tau
    \label{eq:noise-char}
\end{equation}

Given measurements of $x_t$, $\eta_t$, $u_t$, $x_{t+1}$, and $\eta_{t+1}$, we can regress a least-squares approximation for $A$:
\begin{equation}
    A^\mathrm{o}=\mathrm{E}\left[ \phi_{t+1} x_t^* \right]\left(\mathrm{E}\left[x_t^* x_t^{*\intercal}+\epsilon_t \epsilon_t^\intercal \right]\right)^{-1}
\end{equation}
where $x_t^*\triangleq \left(x_t^\intercal, \eta_t^\intercal, u_t^\intercal\right)^\intercal$ and $\epsilon_t\triangleq \left( \epsilon_{x,t}^\intercal, \epsilon_{\eta,t}^\intercal \right)^\intercal$. Even though $A^\mathrm{o}$ does not converge to the optimal noiseless estimate, this problem is well treated in the DMD literature, where total least squares estimation is proposed as an appropriate solution \cite{dmd-noise}.

This issue becomes more severe when using virtual measurements instead of physical ones, as is the case for KSOS. If $\phi_t$ is computed from a noisy state measurement, i.e. $\phi_t\triangleq \phi(\tilde x_t)$ instead of $\phi_t\triangleq \phi(x_t)$, the noise characteristics in equation (\ref{eq:noise-char}) are no longer necessarily zero, but will instead depend on the particular functional forms of $\phi$. Because the states are correlated through the dynamics and the noise is state-dependent, traditional solutions to regress $A^\mathrm{o}$ will not converge to an optimal noiseless estimate.





\end{document}